\documentclass[twocolumn,english,pra,showpacs]{revtex4-1}
\usepackage[T1]{fontenc}
\usepackage{babel}
\usepackage{amsmath}
\usepackage{amssymb}
\usepackage{graphicx}
\usepackage{esint}
\usepackage[unicode=true]{hyperref}
\hypersetup{
 colorlinks,linkcolor=blue,citecolor=blue}

\begin{document}

\title{F\"orster resonance energy transfer, absorption and emission spectra
in multichromophoric systems: I. Cumulant expansions}

\author{Jian Ma}

\author{Jianshu Cao}

\email{jianshu@mit.edu}

\address{Department of Chemistry, Massachusetts Institute of Technology, Cambridge,
MA 02139, USA}
\begin{abstract}
We study the F\"orster resonant energy transfer (FRET) rate in multichromophoric
systems. The multichromophoric FRET rate is determined by the overlap
integral of the donor's emission and acceptor's absorption spectra,
which are obtained via 2nd-order cumulant expansion techniques developed
in this work. We calculate the spectra and multichromophoric FRET
rate for both localized and delocalized systems. (i) The role of the
initial entanglement between the donor and its bath is found to be
crucial in both the emission spectrum and the multichromophoric FRET
rate. (ii) The absorption spectra obtained by the cumulant expansion
method are quite close to the exact one for both localized and delocalized
systems, even when the system-bath coupling is far from the perturbative
regime. (iii) For the emission spectra, the cumulant expansion can
give very good results for the localized system, but fail to obtain
reliable spectra of the high excitations of a delocalized system,
when the system-bath coupling is large and the thermal energy is small.
(iv) Even though, the multichromophoric FRET rate is good enough since
it is determined by the overlap integral of the spectra. 
\end{abstract}



\maketitle

\section{Introduction}

\textit{Background.} Excitonic energy transfer (EET) \cite{Agranovich,Andrews,May,Micha,Foster}
attracts extensive interest in many subjects. It is a fundamental
problem in various physical and chemical processes \cite{Blankenship,renger_ultrafast_2001,novoderezhkin_physical_2010,varnavski_investigations_2002,nguyen_control_2000,bopp_dynamics_1999,oijen_unraveling_1999,ye_excitonic_2012}.
In general, the efficiency of the EET can be well quantified by the
F{ö}rster resonance energy transfer (FRET) theory \cite{Foster,Agranovich,Andrews}
under the following two conditions. (a) The system can be treated
as two parts: the donor and acceptor, and the coupling between them
is much weaker then the system-bath coupling, i.e., the transfer is
usually incoherent. (b) Both acceptor and donor can be treated as
point dipoles.

However, the FRET theory is problematic in real systems such as light-harvesting
complexes LH1/LH2 \cite{bopp_dynamics_1999,oijen_unraveling_1999},
dendrimers \cite{varnavski_investigations_2002,nguyen_control_2000},
and conjugated polymers \cite{nguyen_control_2000}. In these systems,
the donor and/or acceptor could have more than one chromophore, and
cannot be treated as point dipoles. Moreover, due to the electronic
couplings $V^{D}$($V^{A}$) within the donor(acceptor), the excitations
are not localized, and their coherent dynamics could be quite important
in the EET process \cite{collini_coherently_2010,engel_evidence_2007,mercer_instantaneous_2009,lee_coherence_2007,Ishizaki:2010fk},
which was shown recently in two-dimensional electronic spectroscopy
experiments \cite{abramavicius_coherent_2009,cho_coherent_2008,jonas_optical_2003}.
The energy transfer rate is significantly underestimated by the FRET
theory, such as in the LH2 complex \cite{herek_b800-->b850_2000,hu_photosynthetic_2002,beljonne_beyond_2009}.
Therefore, the multichromophoric FRET theory was developed \cite{sumi_theory_1999,scholes_long-range_2003,jang_multichromophoric_2004}
to solve this problem. 

It should be noted that, under some experimental conditions, even
the extension from the FRET to multichromophoric FRET is not enough,
since they are 2nd-order theory with respect to the donor-acceptor
coupling $H_{c}$. Nontrivial quantum effects such as multi-site quantum
coherence and solvent-controlled transfer can be seen in higher order
corrections \cite{Wu2013_HighOrder}.

Similar to its single chromophoric counterpart, the multichromophoric
FRET rate is determined by an overlap integral between the donor's
emission and acceptor's absorption spectra. The spectra are broadened
and shifted due to the environment, which is believed to play a critical
role in the EET process of light-harvesting complexes, which could
achieve the order of a few picoseconds \cite{mennucci_role_2011}.
Unlike the case in the FRET theory, where the spectra can be obtained
exactly for an environment with Gaussian fluctuations \cite{Mukamelve},
the spectra in the multichromophoric FRET theory are more involved,
especially the emission spectrum.

\subsection{Outline of this paper}

\textit{Absorption and Emission spectra.} In the calculation of the
multichromophoric FRET rate, the absorption spectrum is relatively
easier to obtain since the initial state is factorized. The emission
spectrum is much more complicated due to the initial system-bath coupling,
which displaces the bath away from its equilibrium. This displacement
will affect the subsequent dynamics, which is reflected by a complex-time
correlation function. 

\textit{Donor-bath entanglement.} The influence of the initial entanglement
or correlation between the donor and bath is widely noticed but lack
of systemic study, partially due to the difficulties in numerical
techniques. Moreover, this problem does not happen in the monomer
case, where the system is only one-dimensional. In this work, we find
that the donor-bath entanglement plays a crucial role in both the
emission spectrum and multichromophoric FRET rate. Exact numerical
comparisons show the failure of two different factorization approaches
for both localized and delocalized systems.

\textit{Full 2nd-order cumulant expansion.} The primary goal of this
series of papers is to develop analytical and numerical techniques
to compute the multichromophoric FRET rate and spectra. Nonperturbative
numerical methods such as stochastic path integrals \cite{Moix2013_PI_FRET}
and hierarchy equation of motions (HEOM) \cite{Tanimura2006fk,Tanimura1993,Ishizaki:2010fk,Ishizaki:2009vn,Ishizaki:2009zr,chen_optical_2009,Struempfer:2011uq,jing_equilibrium_2013}
can give benchmarks, however, for small systems only due to the limitation
of computing powers. Perturbative methods \cite{sumi_theory_1999,jang_single_2003,renger_relation_2002,banchi_analytical_2013,maier_charge_2011,Lee:2012we,Yang2005,zhang_exciton-migration_1998,kolli_electronic_2011,novoderezhkin_energy-transfer_2004,cho_exciton_2005,yang_influence_2002,ohta_ultrafast_2001,schroder_calculation_2006}
are efficient for larger systems but only in some specific parameter
regimes. For example, in the weak system-bath coupling regime, the
EET was generally studied by using Green's function \cite{sumi_theory_1999},
2nd-order time-convolution (TC2) \cite{jang_multichromophoric_2004,jang_single_2003}
and time-convolution-less (TCL2) master equations \cite{renger_relation_2002,banchi_analytical_2013}

In this paper, we demonstrate the difficulties and problems in the
multichromophoric FRET spectra, and focus on a perturbation approach
based on the 2nd-order cumulant expansion. Here, the cumulant expansion
is performed on the full system-bath coupling Hamiltonian $H_{sb}$
in both the real- and imaginary-time domains. Therefore, the absorption
and emission spectra are expressed in full 2nd-order cumulant expansions
(FCE), which can reduce to the exact results in the monomer case.
As previously shown in the calculation of vibrational spectra \cite{Yang2005a,schroder_calculation_2006},
factorization of the FCE leads to further approximations that are
easy to evaluate analytically. In the exciton bases $H_{sb}$ will
have off-diagonal terms. If the off-diagonal part $H_{sb}^{od}$ is
neglected, the FCE reduces to the inverse participation ratio (IPR)
\cite{cho_exciton_2005,Cleary2013_IPR} method. The IPR method can
be improved by treating the off-diagonal elements perturbatively
\cite{Yang2005a,schroder_calculation_2006}, and
here we call this method the off-diagonal cumulant expansion (OCE).
The advantage of the FCE over the OCE and IPR methods can be seen
in a highly delocalized case, where the omission or perturbation treatment
of the off-diagonal coupling is unreliable. For the absorption spectrum,
the FCE is the same as the TCL2 in formalism, unlike the TC2 method
which cannot reduce to the monomer case. For the emission spectrum,
both the TCL2 and TC2 methods cannot reduce to the monomer case. The
TCL2 method needs the help of a detailed balance identity to overcome
this problem \cite{banchi_analytical_2013}. However, the FCE is more
straightforward and can reduce to the monomer case naturally.

\textit{Reliability.} We use the FCE method to calculate the spectra
and the multichromophoric FRET rate for both localized and delocalized
systems. Firstly, for both systems, the FCE method is quite reliable
in the absorption spectrum, since there is no population dynamics
while the coherence decays so fast as the increase of the system-bath
coupling.

For the emission spectrum, the FCE method is also quite reliable when
the free excitations of the donor is localized, since the FCE is exact
for monomers. However, if the donor's free excitations are delocalized,
the perturbation of the imaginary-time part is unreliable when the
system's energy gap is larger than the thermal energy. In this case,
the emission spectrum is still reliable if it is determined by the
lower excited states. The multichromophoric FRET rates of the two
systems are still close to the exact one since the rate is determined
by the spectra overlap.

This paper is organized as follows. In Sec.~II we give the physical
model of a multichromophoric system and introduce the multichromophoric
FRET theory. The role of the initial entanglement is showed by using
the HEOM method. In Sec.~III we derive the absorption and emission
spectra by using 2nd-order cumulant expansion techniques. We find
that the spectra formula can be further simplified when the system
has translational symmetry. Then we calculate the spectra and multichromophoric
FRET rate for localized and delocalized systems, and discuss the reliability
of the cumulant expansion method.

\subsection{Outline of the forth-coming papers}

We find that the limitation of the FCE method, as well as many other
traditional perturbation methods, lies in the emission spectrum of
a delocalized system in the low-temperature and large system-bath
coupling regimes. To overcome this problem, several new methods are
developed in our group and will appear as a sequel.

(a) For real systems such as the LH2, the energy gaps of the first
excitations are comparable to both the thermal energy and the system-bath
coupling. All the traditional perturbation methods cannot give reliable
emission spectrum and the multichromophoric FRET rate. For such systems,
the treatments of the complex-time system-bath correlation will determine
the reliability of the emission spectrum and the multichromophoric
FRET rate. In our Paper II \cite{Ma2013_new_cumulant}, we develop
a hybrid cumulant expansion method, which uses the imaginary-time
path integrals to obtain the exact reduced density matrix of the donor,
from which the displacements of the bath operators can be extracted
more precisely. Using this method, we can give reliable emission spectrum
and multichromophoric FRET rate of the LH2 system \cite{Moix2013_PI_FRET}.

(b) If the system-bath coupling is dominant, even when the donor's
free excitations are delocalized, perturbation should be made on the
donor's off-diagonal coupling $V$ but not the system-bath coupling.
This O($V^{2}$) expansion is developed in our paper III \cite{Cleary}.

(c) Furthermore, to overcome the problems of the HEOM in calculating
large system and low temperature conditions, in our Paper IV \cite{Moix2013_PI_FRET}
we implement a complex-time stochastic PI method, which will be our
benchmarks.

\section{Multichromophoric FRET Theory}

\subsection{Model Hamiltonian}

The multichromophoric FRET theory describes the resonant energy transfer
between a donor (D) and an acceptor (A) in an multichromophoric system,
described by the Hamiltonian 
\begin{equation}
H=H_{t}^{D}+H_{t}^{A}+H_{c},
\end{equation}
where $H_{c}$ is the dipole-dipole coupling between the donor and
acceptor, and $H_{t}^{D(A)}$ is the total Hamiltonian of the donor
(acceptor) and its bath, 
\begin{eqnarray}
H_{t}^{D} & = & H_{s}^{D}+H_{sb}^{D}+\mathbb{I}_{s}^{D}H_{b},\nonumber \\
H_{t}^{A} & = & H_{s}^{A}+H_{sb}^{A}+\mathbb{I}_{s}^{A}H_{b}.
\end{eqnarray}
We first explain the donor's part. The free Hamiltonian of the donor
is 
\begin{eqnarray}
H_{s}^{D}\!\!\! & =\!\!\! & \sum_{m=1}^{N_{D}}\!\left(\epsilon_{m}^{D}\!+\lambda_{m}^{D}\right)\!\!|D_{m}\rangle\langle D_{m}|\!+\!\sum_{m\neq n}^{N_{D}}\! V_{mn}^{D}|D_{m}\rangle\langle D_{n}|\label{eq:Donors_Hamiltonian}
\end{eqnarray}
where $\epsilon_{m}^{D}$ is the excitation energy of the donor's
$m$th site, and $\lambda_{m}^{D}$ is the reorganization energy induced
by the interaction between the bath and the donor's $m$th chromophore.
$V_{mn}^{D}$ is the coupling between sites $m$ and $n$. In the
multichromophoric FRET theory, we focus on the single excitation case
and thus $|D_{m}\rangle$ represents the state that the total multichromophoric
system is excited only at the donor's $m$th site, all the other sites
(including the acceptor's) are in their ground states, i.e., 
\begin{equation}
|D_{m}\rangle=|0,\dots1_{m},\dots,0\rangle_{D}|0\dots0\rangle_{A}.
\end{equation}
The identity operator $\mathbb{I}_{s}^{D}$ is given by 
\begin{equation}
\mathbb{I}_{s}^{D}=\sum_{m=1}^{N_{D}}|D_{m}\rangle\langle D_{m}|.
\end{equation}

In this work, the baths that couple to different chromophores are
independent. Therefore, it is convenient to write the free Hamiltonian
of the bath as $H_{b}=H_{b}^{D}+H_{b}^{A}$, and then we can write
the total Hamiltonians as 
\begin{align}
H_{t}^{D} & =H^{D}+\mathbb{I}_{s}^{D}H_{b}^{A},\nonumber \\
H_{t}^{A} & =H^{A}+\mathbb{I}_{s}^{A}H_{b}^{D},
\end{align}
which will be used in the following content. The correlated
bath was studied in Ref.~\cite{Wu2010,Struempfer:2011uq}. The bath
is usually modeled by a set of harmonic oscillators, 
\begin{equation}
H_{b}^{D}=\sum_{m=1}^{N_{D}}\sum_{k}\hbar\omega_{m,k}^{D}b_{m,k}^{D\dagger}b_{m,k}^{D},
\end{equation}
where $\omega_{m,k}^{D}$ is the frequency of the $k$th mode of the
bath that coupled with the $m$th site of the donor(acceptor). The
excitation states couple with the harmonic bath linearly as 
\begin{eqnarray}
H_{sb}^{D} & = & \sum_{m=1}^{N_{D}}\hat{B}_{m}^{D}|D_{m}\rangle\langle D_{m}|,\label{H_sb}
\end{eqnarray}
where the bath operators are given by 
\begin{eqnarray}
\hat{B}_{m}^{D} & = & \sum_{k}g_{m,k}^{D}\left(b_{m,k}^{D\dagger}+b_{m,k}^{D}\right).
\end{eqnarray}
The relation between the coupling strengths $g_{m,k}^{D}$ and the
reorganization energy is $\lambda_{m}^{D}\equiv\sum_{k}g_{m,k}^{2}/\omega_{m,k}$.

The acceptor's Hamiltonians $H_{s}^{A}$, $H_{b}^{A}$ and $H_{sb}^{A}$
are obtained by replacing the notation $D$ with $A$ in the above
discussion.

The dipole-dipole interaction between the donor and acceptor is given
by 
\begin{equation}
H_{c}=\sum_{m=1}^{N_{D}}\sum_{n=1}^{N_{A}}J_{mn}\left(|D_{m}\rangle\langle A_{n}|+\text{\ensuremath{|A_{n}\rangle\langle D_{m}|}}\right),
\end{equation}
where the couplings $J_{mn}$ are treated perturbatively in the multichromophoric
FRET theory.

\subsection{Gold-rule formulation of the multichromophoric FRET rate}

Before the study of the multichromophoric FRET rate $k$, we should
understand the time scales in the energy transfer process. The multichromophoric
FRET theory describes the incoherent transfer of excitations from
a donor to an acceptor. This transfer happens after the donor is excited
to its first excitation states. In general, the donor's initial excitations
will relax to an equilibrium state with its bath in a time scale that
is much shorter than the excitation transfer time $1/k$. Therefore,
the initial condition of the multichromophoric FRET process can be
considered as an equilibrium state of the donor and its bath. On the
other hand, the lifetime of the first excitations are usually much
longer than the excitation transfer time, and thus the ground state
is not involved in the multichromophoric FRET theory.

Based on the above conditions, the multichromophoric FRET rate can
be derived straightforwardly from the Fermi's golden rule \cite{sumi_theory_1999},
\begin{equation}
k=2\pi\sum_{\mu\nu}P_{\nu}^{D}\left|\langle\Psi_{\nu}^{D}|H_{c}|\Phi_{\mu}^{A}\rangle\right|^{2}\delta\left(E_{\nu}^{D}-E_{\mu}^{A}\right),\label{eq:MC-FRET_gr}
\end{equation}
where $|\Psi_{\nu}^{D}\rangle$ ($|\Phi_{\mu}^{A}\rangle$) and $E_{\nu}^{D}$
($E_{\mu}^{A}$) are the eigenstates and eigenenergies of $H_{t}^{D}$
($H_{t}^{A}$), which include the degrees of freedom of both the system
and bath. $P_{\nu}^{D}$ is obtained from 
\begin{equation}
\rho^{D}=\rho_{e}^{D}\rho_{b}^{A}=\sum_{\nu}P_{\nu}^{D}|\Psi_{\nu}^{D}\rangle\langle\Psi_{\nu}^{D}|,
\end{equation}
where 
\begin{equation}
\rho_{e}^{D}=\frac{e^{-\beta H^{D}}}{\text{tr}e^{-\beta H^{D}}},\;\rho_{b}^{A}=\frac{e^{-\beta H_{b}^{A}}}{\text{tr}e^{-\beta H_{b}^{A}}},
\end{equation}
and $\beta=k_{B}T$, with $k_{B}$ the Boltzmann's constant and $T$
the temperature.

Starting from Eq.~(\ref{eq:MC-FRET_gr}), the multichromophoric FRET
rate can be derived as 
\begin{eqnarray}
k= &  & \sum_{m,n}\sum_{m'n'}J_{mn}J_{m'n'}\int_{-\infty}^{\infty}dt\,\text{tr}_{b}\big\{ e^{iH_{t}^{D}t}\rho_{e}^{D}\rho_{b}^{A}\nonumber \\
 &  & \times|D_{m'}\rangle\langle A_{n'}|e^{-iH_{t}^{A}t}|A_{n}\rangle\langle D_{m}|\big\},
\end{eqnarray}
where the degrees of freedom of the acceptor and donor can be treated
separately as 
\begin{align}
k & =\sum_{m,n}\sum_{m'n'}J_{mn}J_{m'n'}\int_{-\infty}^{\infty}dt\,\nonumber \\
 & \times\text{tr}_{bA}\left[\rho_{b}^{A}e^{iH_{b}^{A}t}\langle A_{n'}|e^{-iH^{A}t}|A_{n}\rangle\right]\nonumber \\
 & \times\text{tr}_{bD}\left[\langle D_{m}|e^{iH^{D}t}\rho_{e}^{D}|D_{m'}\rangle e^{-iH_{b}^{D}t}\right].
\end{align}
Now we can define two matrices 
\begin{eqnarray}
\mathbf{I}^{A}\left(t\right) & = & \text{tr}_{b}\left(e^{-iH^{A}t}\rho_{b}^{A}e^{iH_{b}^{A}t}\right),\label{eq:IA}\\
\mathbf{E}^{D}\left(t\right) & = & \text{tr}_{b}\left(e^{iH^{D}t}\rho_{e}^{D}e^{-iH_{b}^{D}t}\right),\label{eq:ED}
\end{eqnarray}
and the multichromophoric FRET rate can be expressed as 
\begin{eqnarray}
k & = & \int_{-\infty}^{\infty}dt\,\text{tr}\left[\mathbf{J}^{T}\mathbf{E}^{D}\left(t\right)\mathbf{J}\mathbf{I}^{A}\left(t\right)\right],
\end{eqnarray}
where 
\begin{equation}
\mathbf{J}=\sum J_{mn}|D_{m}\rangle\langle A_{n}|.
\end{equation}
It is important to notice that the density matrix in $\mathbf{I}^{A}\left(t\right)$
is the thermal equilibrium state of the acceptor's bath. The donor
is assumed to be in its excited equilibrium state $\rho_{e}^{D}$.
This non-factorized initial state brings technique difficulties in
the calculation of emission spectrum, especially when the system-bath
coupling is strong. The primary goal of this series of papers is to
develop analytical and numerical techniques to compute the donor-bath
correlations.

The absorption and emission spectra are given by 
\begin{eqnarray}
\mathbf{I}^{A}\left(\omega\right) & = & \int_{-\infty}^{\infty}dt\, e^{i\omega t}\mathbf{I}^{A}\left(t\right),\nonumber \\
\mathbf{E}^{D}\left(\omega\right) & = & \int_{-\infty}^{\infty}dt\, e^{-i\omega t}\mathbf{E}^{D}\left(t\right),\label{eq:MC-FRET spectra}
\end{eqnarray}
and thus the multichromophoric FRET rate can also be written as \cite{sumi_theory_1999,jang_multichromophoric_2004}
\begin{equation}
k=\frac{1}{2\pi}\int_{-\infty}^{\infty}d\omega\,\text{tr}\left[\mathbf{J}^{T}\mathbf{E}^{D}\left(\omega\right)\mathbf{J}\mathbf{I}^{A}\left(\omega\right)\right].\label{eq:MC-FRET_formula}
\end{equation}
From the above formula, the rate $k$ is determined by the donor-acceptor
coupling $\mathbf{J}$, and the overlap integral of the acceptor's
absorption spectrum $\mathbf{I}_{nn'}^{A}\left(\omega\right)$ and
the donor's emission spectrum $\mathbf{E}_{m'm}^{D}\left(\omega\right)$.
The influences of the system-bath coupling on the transfer rate are
reflected by the spectra in their widths and positions, which are
determined by the relaxation dynamics and reorganization energies,
respectively. Therefore, the main problem here is to calculate the
spectra. We should note that the spectra (\ref{eq:MC-FRET spectra})
in the multichromophoric FRET rate do not depend on the system's local
dipoles. Actually, the commonly studied far-field spectra $I_{\text{f}}^{A}\left(\omega\right)$
and $E_{\text{f}}^{D}\left(\omega\right)$ can be obtained as 
\begin{align}
I_{\text{f}}^{A}\left(\omega\right) & =\sum_{m,n}\left(\hat{\epsilon}\cdot\vec{\mu}_{m}^{A}\right)\left(\hat{\epsilon}\cdot\vec{\mu}_{n}^{A}\right)\, I_{mn}^{A}\left(\omega\right),\nonumber \\
E_{\text{f}}^{D}\left(\omega\right) & =\sum_{m,n}\left(\hat{\epsilon}\cdot\vec{\mu}_{m}^{D}\right)\left(\hat{\epsilon}\cdot\vec{\mu}_{n}^{D}\right)\, E_{mn}^{D}\left(\omega\right),
\end{align}
where $\hat{\epsilon}$ is the polarization of the light and $\vec{\mu}_{i}$
denote the local dipole operators. 

In this work, for the sake of simplicity, the donor-acceptor coupling
is $J_{mn}=J$. Therefore, the multichromophoric FRET rate formula
(\ref{eq:MC-FRET_formula}) can be simplified as 
\begin{equation}
k=\frac{J^{2}}{2\pi\hbar^{2}}\int_{-\infty}^{\infty}d\omega E^{D}\left(\omega\right)I^{A}\left(\omega\right),
\end{equation}
where 
\begin{eqnarray}
E^{D}\left(\omega\right) & = & \sum_{m,n}E_{mn}^{D}\left(\omega\right),\nonumber \\
I^{A}\left(\omega\right) & = & \sum_{m,n}I_{mn}^{A}\left(\omega\right).
\end{eqnarray}

\section{Effects of donor-bath entanglement}

In the multichromophoric FRET theory, the donor is first excited to
its single-excitation subspace, which becomes equilibrium with its
bath in a time scale that is negligibly small as compared with the
EET time. Therefore, the initial state is an equilibrium state of
the donor and its bath, as shown in Eq.~(\ref{eq:ED}). In this case,
the donor and its bath are usually correlated or entangled due to
their interaction, which is characterized by the reorganization energy
$\lambda$.

When $\lambda$ is smaller than the system's energy scale, or the
bath correlation time is negligibly small (e.g. in the high-temperature
limit), the Born approximation is employed and a master equation is
obtained. However, when the system-bath interaction is larger than
the other energy scales, the Born approximation is invalid, and the
system and bath are non-factorized during the entire multichromophoric
FRET process. To our knowledge, only a few methods can treat the system-bath
correlation exactly. Here we first use the HEOM method to show the
crucial role of the donor-bath entanglement. The 2nd-order correction
of the initial state was studied in Sec.~IV B.

In this work, the donor and acceptor each consists of two chromophores.
Since the entanglement relies on the properties of the system, we
consider two limiting cases. In the Case I, the system is localized,
and its Hamiltonian is (in the unit of $\text{cm}^{-1}$) 
\begin{equation}
H_{s}^{D1}=\begin{pmatrix}250 & 20\\
20 & 150
\end{pmatrix},\; H_{s}^{A1}=\begin{pmatrix}100 & 20\\
20 & 0
\end{pmatrix},\label{eq:Ham_Local}
\end{equation}
where the ratio of the excitation energy difference $\Delta=E_{2}-E_{1}$
and the inter-chromophore coupling $V$ is $\Delta/V=5$. In the Case
II, the system is delocalized ($\Delta/V$=0.2),

\begin{equation}
H_{s}^{D2}=\begin{pmatrix}200 & 100\\
100 & 180
\end{pmatrix},\; H_{s}^{A2}=\begin{pmatrix}100 & 100\\
100 & 80
\end{pmatrix}.\label{eq:Ham_DeLocal}
\end{equation}
The influence of the bath on the system dynamics is determined by
the system-bath coupling spectrum, here we choose the Drude spectrum
\begin{equation}
J\left(\omega\right)=\frac{2\lambda\omega\gamma}{\omega^{2}+\gamma^{2}},\label{eq:Drude_Spectrum}
\end{equation}
where $\lambda$ is the reorganization energy and $\gamma$ is the
cutoff frequency of the bath. For the sake of simplicity, the reorganization
energies are the same for each site. The donor-acceptor coupling is
$J_{mn}=J=10\,\text{cm}^{-1}$.

\begin{figure}
\includegraphics[width=1\columnwidth]{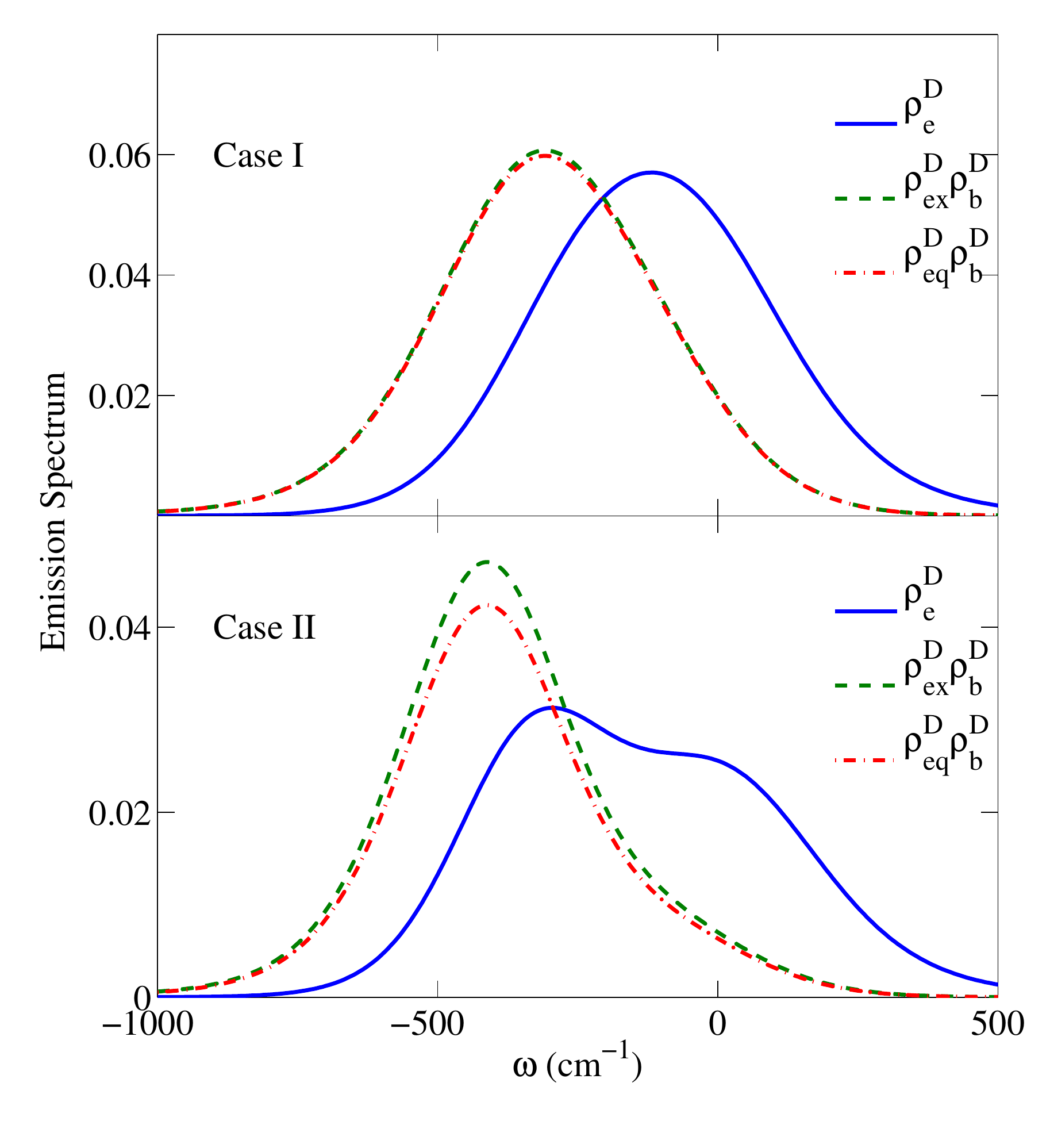}

\protect\caption{Comparison of emission spectra for different initial state in localized
(Case I) and delocalized (Case II) systems. The reorganization energy
$\lambda=100\,\text{cm}^{-1}$, the cutoff frequency $\gamma=10\,\text{ps}^{-1}$,
and the temperature $T=300\,\text{K}$.}

\label{fig:comp_Emission_spectra} 
\end{figure}

To study the effects of the initial entanglement, we consider three
different treatments of the initial state:

(i) The initial state is obtained exactly by using HEOM. In this case,
the system and bath first evolved to equilibrium, and the correlation
information between system and bath is carried by the auxiliary fields
of the HEOM. Then the system and bath evolve according to Eq.~(\ref{eq:ED}).
The emission spectrum obtained in this case is exact \cite{jing_equilibrium_2013}.

\begin{figure}
\includegraphics[width=1\columnwidth]{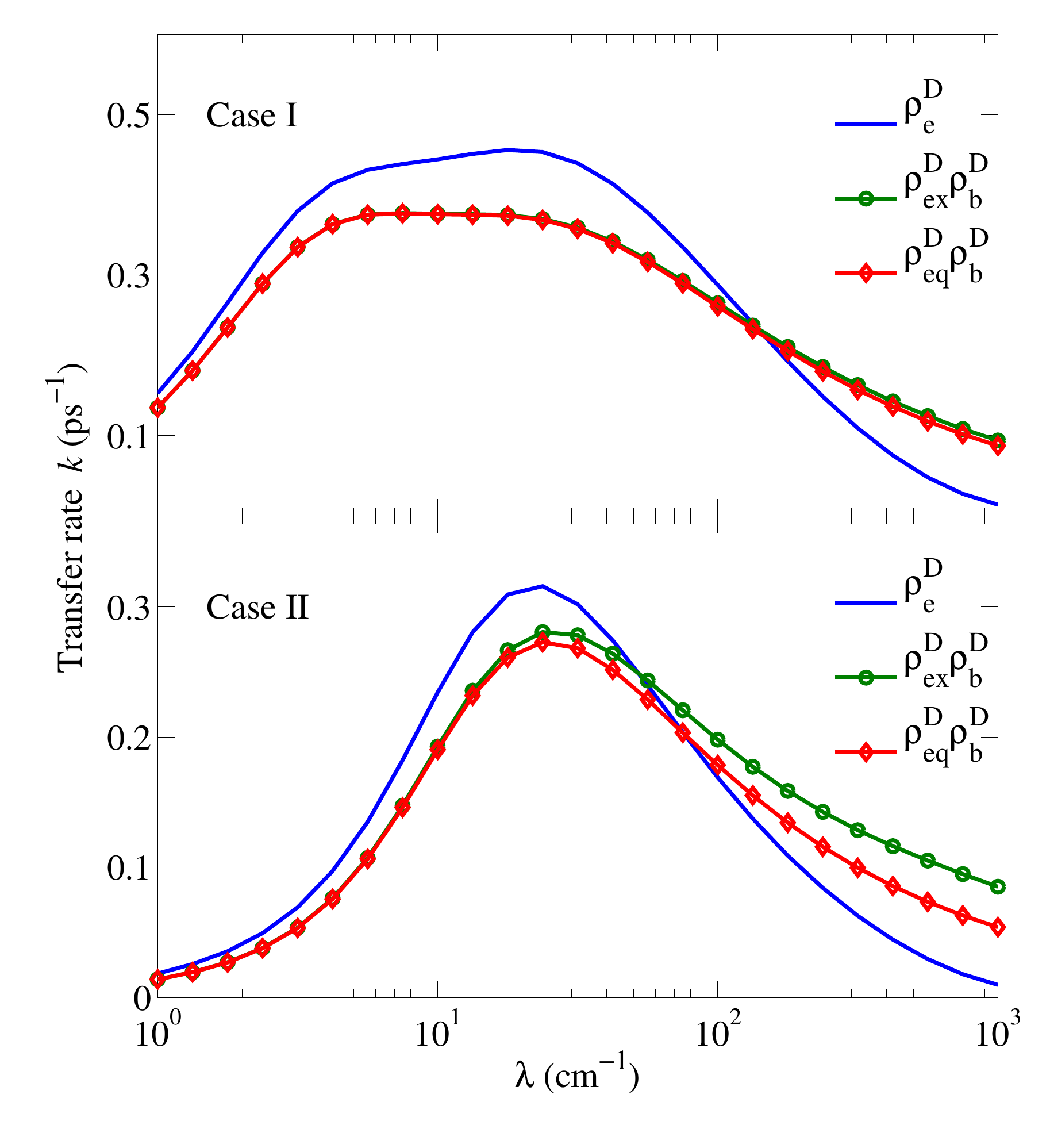}

\protect\caption{Comparison of multichromophoric FRET rate as a function of reorganization
energy $\lambda$ from $1\,\text{cm}^{-1}$ to $1000\,\text{cm}^{1}$
for different initial states. All the other parameters are the same
as in Fig.~\ref{fig:comp_Emission_spectra}.}

\label{fig:comp_MC-FRET} 
\end{figure}

(ii) The initial state is factorized, but the system's reduced density
matrix is exact, 
\begin{equation}
\rho\left(0\right)=\rho_{\text{ex}}^{D}\rho_{b}^{D},
\end{equation}
where $\rho_{\text{ex}}^{D}$ is the exact reduced density matrix
of the donor, and $\rho_{b}^{D}=\exp\left(-\beta H_{b}^{D}\right)/Z_{b}^{D}$
is the equilibrium state of the bath. In this case, we use the HEOM
method to obtain the exact reduced density matrix, then all the auxiliary
fields are reset to be zeros. Thus the correlation between system
and bath is turned off. Alternatively, we can obtain the reduced density
matrix directly using the imaginary-time path integral method \cite{Moix2012}. 

(iii) The initial state is also factorized as 
\begin{equation}
\rho\left(0\right)=\rho_{\text{eq}}^{D}\rho_{b}^{D},
\end{equation}
however, $\rho_{\text{eq}}^{D}=\exp\left(-\beta H_{s}^{D}\right)/Z_{s}^{D}$
is the thermal equilibrium state of donor. This is also the initial
state commonly used in master equation method.

Comparison of emission spectra $E^{D}\left(\omega\right)$ for different
initial states are shown in Fig.~\ref{fig:comp_Emission_spectra}.
Because the reorganization energy $\lambda=100\,\text{cm}^{-1}$ is
comparable to the excitation energies, thus the separable approximation
is not reliable. For the localized system (Case I), the spectra obtained
by separable approximation are shifted due to the ignore of reorganization.
The delocalized system (Case II) is more interesting, where the double-peak
structure disappears in the approximate spectra. The entanglement
effect in the delocalized case is more notable than in the localized
one. We consider a limiting case, the system is fully localized, i.e.,
the intermolecular coupling $V=0$, thus the chromophores are independent,
and there is no entanglement between the donor and its bath. Actually,
in this case the initial state can be written as 
\begin{equation}
\rho\left(0\right)=\rho_{\text{eq}}^{D}\tilde{\rho}_{b}^{D},
\end{equation}
where $\tilde{\rho}_{b}^{D}$ is the equilibrium state of the displaced
bath.

The effects of initial entanglement on multichromophoric FRET rate
is shown in Fig.~\ref{fig:comp_MC-FRET}. For both the localized
and delocalized cases, the factorized initial state approximation
breaks down rapidly with the increase of system-bath coupling. It
is interesting to note that, in the delocalized case (lower panel)
the multichromophoric FRET rates obtained from $\rho_{\text{ex}}^{D}\rho_{b}^{D}$
and $\rho_{\text{eq}}^{D}\rho_{b}^{D}$ differ dramatically for very
large $\lambda$, which reflects the deviation of $\rho_{\text{ex}}^{D}$
and $\rho_{\text{eq}}^{D}$. As the entanglement plays such a notable
role, we should treat the initial state in a more accurate way, such
as the cumulant expansion method used below.

\section{Second-order Cumulant expansion}

In this section we derive the cumulant expansion formulas of the absorption
and emission spectra. The absorption and emission spectra were studied
by various methods, such as the standard TC2 \cite{jang_single_2003}
and TCL2 \cite{banchi_analytical_2013}. The TC2 method is time-nonlocal,
and it cannot give reliable results even for monomers. The TCL2 method
is time-local. In the monomer case, the absorption spectrum given
by the TCL2 is exact. However, in this framework the TCL2 of the emission
spectrum has an inhomogeneous term that describes the unfactorized
initial states, and it cannot give the exact results in the monomer
limit. A detailed balance identity should be used to overcome this
problem \cite{banchi_analytical_2013}. The cumulant expansion method
shown below is similar to the TCL2 method. They give the same results
for the absorption spectrum. For the emission spectrum, the cumulant
expansion method can reduce to the monomer case directly, without
additional terms.

\subsection{Absorption spectrum -- Full 2nd-order cumulant expansion}

Below, we derive the absorption spectrum via 2nd-order cumulant expansion.
It is convenient to diagonalize the acceptor's Hamiltonian $H_{s}^{A}$
at first, 
\begin{equation}
H_{s}^{A}=\sum_{\mu=1}^{N_{A}}\epsilon_{\mu}^{A}|\mu\rangle\langle\mu|,
\end{equation}
where $\epsilon_{\mu}^{A}$ is the eigenenergy (containing $\lambda$),
\begin{equation}
|\mu\rangle=\sum_{i=1}^{N_{A}}c_{i}^{\mu}|A_{i}^{e}\rangle,
\end{equation}
is the energy eigenstate, and $c_{i}^{\mu}=\langle\mu|A_{i}^{e}\rangle$.
However, in the energy representation the system-bath coupling has
off-diagonal terms, 
\begin{equation}
H_{sb}^{A}=\sum_{\mu,\nu=1}^{N_{A}}\tilde{B}_{\mu\nu}^{A}|\mu\rangle\langle\nu|,\label{eq:H_eB}
\end{equation}
where 
\begin{equation}
\tilde{B}_{\mu\nu}^{A}=\sum_{n=1}^{N_{A}}X_{n}^{\mu\nu}B_{n}^{A}
\end{equation}
and the coefficient $X_{n}^{\mu\nu}=c_{n}^{\mu}c_{n}^{\nu}$. Below,
we perform cumulant expansion with respect to $H_{sb}^{A}$. In Ref.~\cite{Yang2005a,schroder_calculation_2006},
the cumulant expansion was carried out with respect to the off-diagonal
terms of the Hamiltonian $H_{sb}^{A}$ (\ref{eq:H_eB}), which could
yield unreliable results in highly delocalized cases.

\begin{figure}
\includegraphics[width=1\columnwidth]{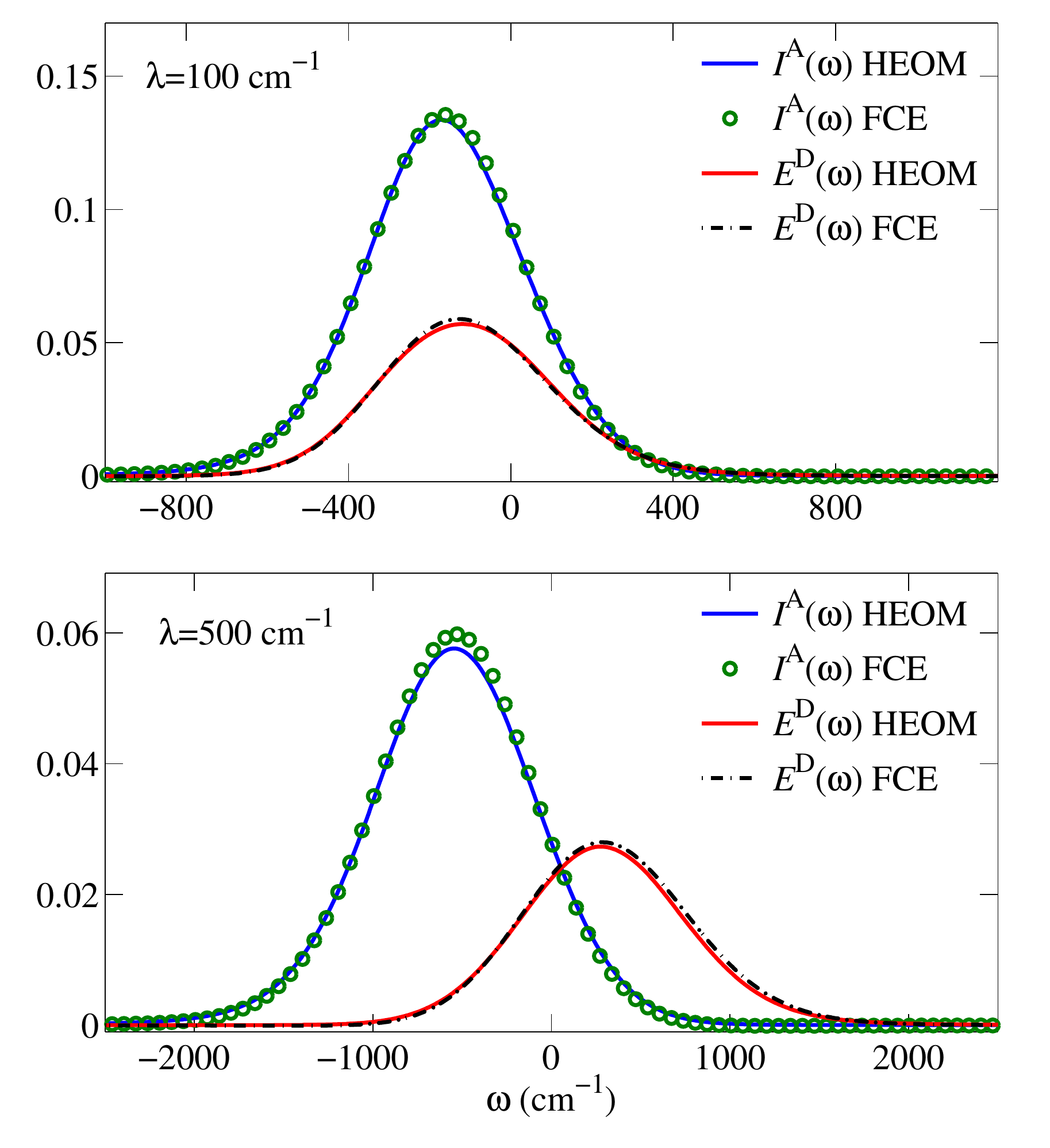}

\protect\caption{Absorption and emission spectra of the localized system {[}Case I
defined in (\ref{eq:Ham_Local}){]}. Results are obtained by using
full 2nd-order cumulant expansion (FCE) and hierarchy equation of
motion (HEOM) methods. The bath parameters are the same as in Fig.~\ref{fig:comp_Emission_spectra}.
Even for a very large reorganization energy $\lambda=500\,\text{cm}^{-1}$,
the FCE results are in very good agreement with the exact one.}

\label{fig:AE_Local} 
\end{figure}

The 2nd-order cumulant of $H_{sb}^{A}$ of Eq.~(\ref{eq:IA}) gives
\begin{eqnarray}
\mathbf{I}^{A}\left(t\right) & \simeq & e^{-iH_{s}^{A}t}e^{-\mathbf{K}\left(t\right)},\label{eq:IA_C}
\end{eqnarray}
where the time-dependent matrix 
\begin{eqnarray}
\mathbf{K}\left(t\right) & = & \!\int_{0}^{t}\!\! dt_{2}\int_{0}^{t_{2}}\!\!\! dt_{1}\text{tr}_{b}\left[H_{sb}^{A}\left(t_{2}\right)H_{sb}^{A}\left(t_{1}\right)\rho_{b}^{A}\right]\nonumber \\
 & = & \sum_{\mu,\nu=1}^{N_{A}}|\mu\rangle\langle\nu|\sum_{\alpha=1}^{N_{A}}\sum_{n=1}^{N_{A}}X_{n}^{\mu\alpha}X_{n}^{\alpha\nu}\nonumber \\
 &  & \times\!\int_{0}^{t}\!\! dt_{2}\int_{0}^{t_{2}}\!\!\! dt_{1}e^{i\omega_{\mu\alpha}t_{2}-i\omega_{\nu\alpha}t_{1}}C_{n}^{B}\left(t_{2}\!-t_{1}\!\right),\label{eq:K}
\end{eqnarray}
where $\omega_{ij}\equiv\epsilon_{i}-\epsilon_{j}$ and 
\begin{equation}
H_{sb}^{A}\left(t\right)\equiv e^{i\left(H_{s}^{A}+H_{b}^{A}\right)t}H_{sb}^{A}e^{-i\left(H_{s}^{A}+H_{b}^{A}\right)t}.
\end{equation}
The time-correlation function of the bath $C_{n}^{B}\left(t_{2}-t_{1}\right)\equiv\text{tr}_{b}\left[B_{n}^{A}\left(t_{2}\right)B_{n}^{A}\left(t_{1}\right)\rho_{b}^{A}\right]$
is time translational invariant. In a general case if we have a complex
time $\theta=s-i\tau$, the correlation function can be expressed
as 
\begin{eqnarray}
C_{n}^{B}\left(\theta\right) & = & \int_{0}^{\infty}\frac{d\omega}{\pi}J_{n}\left(\omega\right)\frac{\cosh\left[\omega\left(\frac{1}{2}\beta-i\theta\right)\right]}{\sinh\left[\frac{1}{2}\omega\beta\right]},\label{eq:C_B}
\end{eqnarray}
where $J_{n}\left(\omega\right)$ is the coupling spectrum between
the $n$th site and its bath. In this paper, we choose the Drude spectrum
(\ref{eq:Drude_Spectrum}), and assume the reorganization energies
are the same for different baths, i.e., $J\left(\omega\right)=J_{n}\left(\omega\right)$.

\begin{figure}
\includegraphics[width=1\columnwidth]{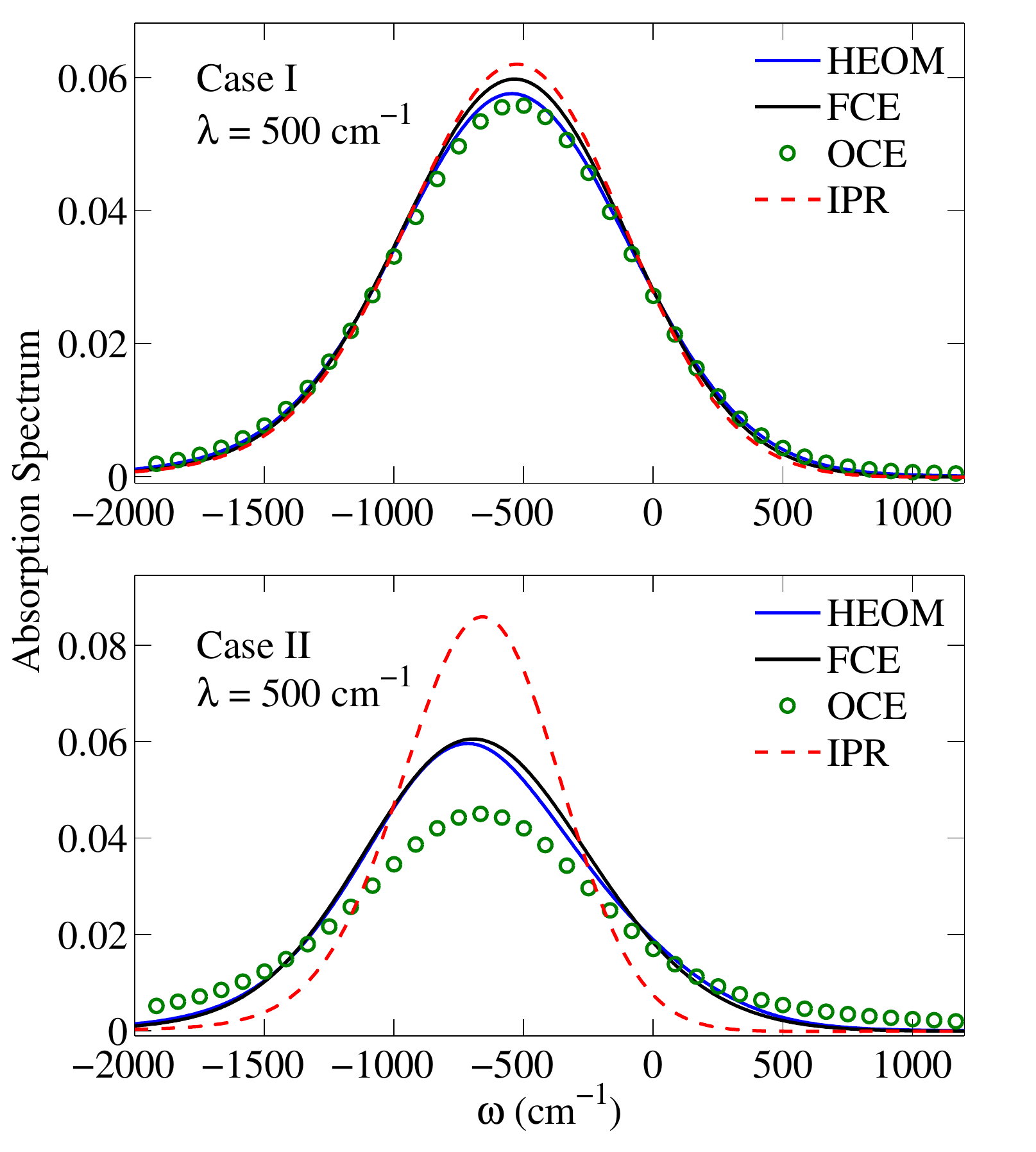}

\protect\caption{Absorption spectra obtained via the hierarchy equation of motion (HEOM),
full 2nd-order cumulant expansion (FCE), off-diagonal cumulant expansion
(OCE) and inverse participation ratio (IPR) methods. The OCE and IPR
methods fail to give reliable results in delocalized case.}

\label{fig:Absorption_Markovian} 
\end{figure}

The calculation of the absorption spectra $I_{A}\left(\omega\right)$
for localized (\ref{eq:Ham_Local}) and delocalized (\ref{eq:Ham_DeLocal})
systems are shown in Figs.~\ref{fig:AE_Local} and \ref{fig:AE_DeLocal},
respectively. We know that if the system is fully localized, i.e.,
$V_{ij}^{D\left(A\right)}=0$, the cumulant expansion method is exact.
Since the Hamiltonian (\ref{eq:Ham_Local}) does not deviate from
the fully localized case very much, the cumulant expansion method
could give very precise results, as shown in Fig.~\ref{fig:AE_Local}.
The cumulant expansion results for the delocalized system are also
in good agreement with the exact spectra obtained by the HEOM method,
even for a very large reorganization energy $\lambda=500\,\text{cm}^{-1}$.

\subsection{Absorption spectrum -- Further approximation of the FCE method}

In general, $\mathbf{K}\left(t\right)$ in Eq.~(\ref{eq:K}) is a
matrix, and Eq.~(\ref{eq:IA_C}) should be evaluated numerically.
In a previous study of vibrational spectra \cite{Yang2005a}, we arrived
at a similar expression and factorized the contribution from the diagonal
and off-diagonal parts of the interaction Hamiltonian

\begin{equation}
H_{sb}^{A}=H_{sb}^{A,d}+H_{sb}^{A,od},
\end{equation}
in the exciton bases, where $H_{sb}^{A,d}$ and $H_{sb}^{A,od}$ are
the diagonal and off-diagonal parts. If we neglect $H_{sb}^{A,od}$
and perform cumulant expansion on the diagonal part $H_{sb}^{A,d}$
only, we will arrive at the IPR method \cite{Cleary2013_IPR,Cleary2013_PNAS},
and the absorption spectrum is calculated as
\begin{align}
\mathbf{I}_{IPR}^{A}\left(t\right) & \simeq\sum_{\mu}|\mu\rangle\langle\mu|\exp\left[-i\epsilon_{\mu}^{A}t-K_{\mu\mu}^{IPR}\left(t\right)\right],
\end{align}
where 
\begin{align}
K_{\mu\mu}^{IPR}\left(t\right) & =\sum_{n=1}^{N_{A}}\left|X_{n}^{\mu\mu}\right|^{2}\int_{0}^{t}\!\! dt_{2}\int_{0}^{t_{2}}\!\!\! dt_{1}C^{B}\left(t_{2}\!-t_{1}\!\right).
\end{align}
We should note that, $K_{\mu\mu}^{IPR}\left(t\right)$ is not the
diagonal part of $\mathbf{K}\left(t\right)$ in Eq.~(\ref{eq:K}).
It is reliable only for localized case, in which the off-diagonal
terms of the system-bath coupling is small. Further improvement can
be made by including the contribution from $H_{sb}^{A,od}$. Actually,
the diagonal part of $\mathbf{K}\left(t\right)$ can be written as
\begin{equation}
K_{\mu\mu}\left(t\right)=K_{\mu\mu}^{IPR}\left(t\right)+\sum_{\alpha}R_{\mu\alpha\alpha\mu}\left(t\right),
\end{equation}
where
\begin{align}
R_{\mu\alpha\alpha\mu}\left(t\right) & =\sum_{\alpha\neq\mu}\sum_{n=1}^{N_{A}}\left(X_{n}^{\mu\alpha}\right)^{2}\nonumber \\
 & \times\int_{0}^{t}\!\! dt_{2}\int_{0}^{t_{2}}\!\!\! dt_{1}e^{i\omega_{\mu\alpha}\left(t_{2}-t_{1}\right)}C\left(t_{2}-t_{1}\right)
\end{align}
reflects the transition from state $\alpha$ to $\mu$, which is induced
by the off-diagonal part of the acceptor-bath interaction Hamiltonian
$H_{sb}^{A,od}$. In the long time limit, $R_{\mu\alpha\alpha\mu}\equiv R_{\mu\alpha\alpha\mu}\left(\infty\right)$
is the population transfer rate. Therefore, the absorption spectrum
can also be given as 
\begin{align}
\mathbf{I}_{OCE}^{A}\left(t\right)\simeq \sum_{\mu}&|\mu\rangle\langle\mu|\exp\bigg[-i\epsilon_{\mu}^{A}t-K_{\mu\mu}^{IPR}\left(t\right)\nonumber \\
 & -\sum_{\alpha\neq\mu}R_{\mu\alpha\alpha\mu}t\bigg],
\end{align}
which is named as the OCE approach. We compare these methods in Fig.~\ref{fig:Absorption_Markovian}.
For localized system, all of them give reliable results. For delocalized
case, since the off-diagonal terms of $H_{sb}^{D(A)}$ are not negligible,
only the FCE can give reliable spectra.

\subsection{Emission spectrum}

The emission spectrum is also derived in the energy representation.
The density matrix $\rho_{e}^{D}$ in Eq.~(\ref{eq:ED}) needs to
be treated carefully. We first consider the partition function $Z_{e}^{D}=\text{tr}\left(e^{-\beta H^{D}}\right)$,
which is obtained by cumulant expansions 
\begin{eqnarray}
Z_{e}^{D} & \simeq & Z_{b}^{D}\text{tr}_{D}\left[e^{-\beta H_{s}^{D}}e^{\mathbf{K}^{II}\left(\beta\right)}\right],
\end{eqnarray}
where the matrix 
\begin{eqnarray}
 & \mathbf{K}^{II}\left(\beta\right) & =\int_{0}^{\beta}\!\! d\tau_{2}\int_{0}^{\tau_{2}}\!\!\! d\tau_{1}\text{tr}_{b}\left[H_{sb}^{D}\!\!\left(-i\tau_{2}\right)H_{sb}^{D}\!\!\left(-i\tau_{1}\right)\rho_{b}^{D}\right]\nonumber \\
 &  & =\sum_{\mu\nu\alpha}\sum_{n}X_{n}^{\mu\alpha}X_{n}^{\alpha\nu}|\mu\rangle\langle\nu|\nonumber \\
 &  & \times\int_{0}^{\beta}d\tau'e^{\omega_{\mu\nu}\tau'}\int_{0}^{\tau'}d\tau\, e^{\omega_{\nu\alpha}\tau}C^{B}\left(-i\tau\right),\label{eq:KII}
\end{eqnarray}
and the imaginary time bath correlation function $C^{B}\left(-i\tau\right)$
is given by Eq.~(\ref{eq:C_B}). After similar algebra used in the
previous section, we obtain 
\begin{equation}
\mathbf{E}_{D}\left(t\right)\simeq\frac{e^{-\left(\beta+it\right)H_{s}^{D}}e^{-\mathbf{K}^{RR}\left(t,\beta\right)+i\mathbf{K}^{RI}\left(t,\beta\right)+\mathbf{K}^{II}\left(\beta\right)}}{\text{tr}_{D}\left[e^{-\beta H_{s}^{D}}e^{\mathbf{K}^{II}\left(\beta\right)}\right]},\label{eq:ED_C}
\end{equation}
where 
\begin{eqnarray}
\mathbf{K}^{RR}\left(t,\beta\right) & = & \!\!\int_{0}^{t}\!\!\! ds_{2}\int_{0}^{s_{2}}\!\!\! ds_{1}\text{tr}_{b}\left[H_{sb}^{D}\!\!\left(s_{2}\!\!-\!\! i\beta\right)H_{sb}^{D}\!\!\left(s_{1}\!\!-\!\! i\beta\right)\rho_{b}^{D}\right]\nonumber \\
 & = & \sum_{\mu\nu\alpha}\sum_{n}X_{n}^{\mu\alpha}X_{n}^{\alpha\nu}|\mu\rangle\langle\nu|e^{\beta\omega_{\mu\nu}}\nonumber \\
 &  & \times\int_{0}^{t}\!\!\! ds'\, e^{i\omega_{\mu\nu}s'}\int_{0}^{s'}\!\!\!\! ds\, e^{i\omega_{\nu\alpha}s}\, C^{B}\left(s\right),\label{eq:KRR}
\end{eqnarray}
and 
\begin{eqnarray}
\mathbf{K}^{RI}\left(t,\beta\right) & = & \!\!\int_{0}^{t}\!\!\! ds\int_{0}^{\beta}\!\!\! d\tau\text{tr}_{b}\left[H_{sb}^{D}\!\!\left(s\!-\! i\beta\right)H_{sb}^{D}\!\!\left(-i\tau\right)\rho_{b}^{D}\right]\nonumber \\
 & = & \sum_{\mu\nu\alpha}\sum_{n}X_{n}^{\mu\alpha}X_{n}^{\alpha\nu}|\mu\rangle\langle\nu|e^{\beta\omega_{\mu\alpha}}\nonumber \\
 & \times & \int_{0}^{t}\!\!\! ds\int_{0}^{\beta}\!\!\!\, d\tau e^{i\omega_{\mu\alpha}s-\omega_{\nu\alpha}\tau}C^{B}\!\left(-s\!-\! i\tau\right).\label{eq:KRI}
\end{eqnarray}
We note that matrices $\mathbf{K}^{RR}\left(t,\beta\right)$ and $\mathbf{K}^{RI}\left(t,\beta\right)$
depend both on time and temperature, which reflect that the dynamics
is affected by the initial system-bath correlation. The explicit forms
of the above matrices for the Drude spectrum are given in Appendix
A.

\begin{figure}
\includegraphics[width=1\columnwidth]{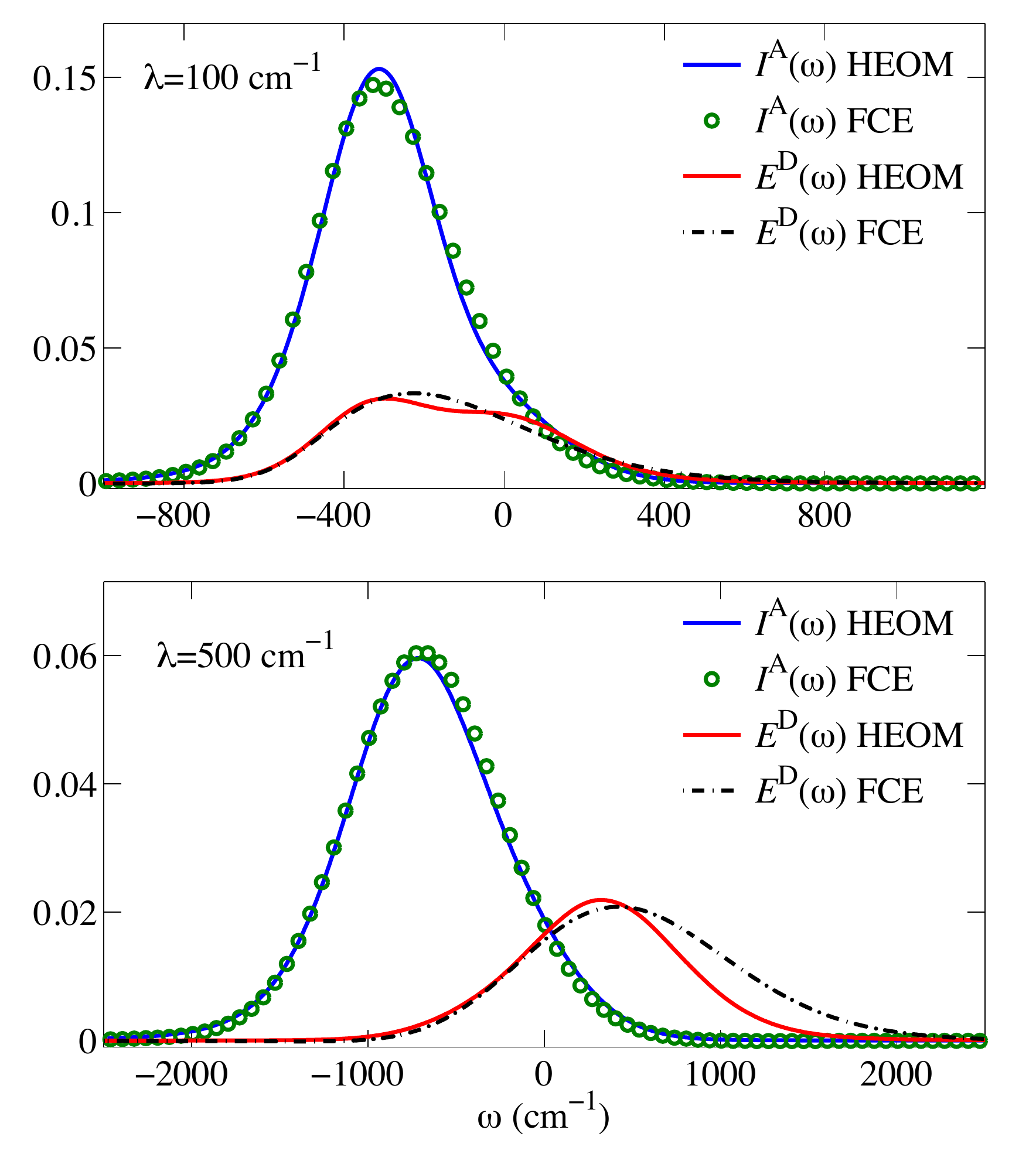}

\protect\caption{Absorption and emission spectra of the delocalized system (\ref{eq:Ham_DeLocal}).
We compare the results of the full 2nd-order cumulant expansion (FCE)
and hierarchy equation of motion (HEOM) methods. The bath parameters
are the same as in Fig.~\ref{fig:comp_Emission_spectra}. }

\label{fig:AE_DeLocal} 
\end{figure}

The comparisons of the emission spectra $E^{D}\left(\omega\right)$
for localized {[}Case I (\ref{eq:Ham_Local}){]} and delocalized {[}Case
II (\ref{eq:Ham_DeLocal}){]} systems are shown in Figs.~\ref{fig:AE_Local}
and \ref{fig:AE_DeLocal}, respectively. For a localized system, the
donor and its bath is just weakly entangled, thus the cumulant expansion
method performs very well, as shown in Fig.~\ref{fig:AE_Local}.
For a delocalized system, the donor and its bath could be strongly
entangled. The emission spectra deviate from the exact one when the
system-bath coupling becomes so strong that the initial state is far
from a factorized state. Actually, perturbative methods are unreliable
in this parameter regime. The reliability of the cumulant expansion
method will be discussed in Sec.~III D.

\subsection{Systems with translational symmetry}

To calculate the emission and absorption spectra, we need to diagonalize
all the matrices $\mathbf{K}$ in every time step according to Eqs.
(\ref{eq:IA_C}) and (\ref{eq:ED_C}). This could be time consuming
if the practical system is large. Fortunately, it can be proved that
the matrices $\mathbf{K}$ are diagonal when a system has translational
symmetry (the reorganization energies are also equal).

All the matrices $\mathbf{K}$ has the factor 
\begin{equation}
\sum_{n}X_{n}^{\mu\alpha}X_{n}^{\alpha\nu}=\sum_{n}\langle\mu|n\rangle\left|\langle n|\alpha\rangle\right|^{2}\langle n|\nu\rangle.
\end{equation}
If the system has translational symmetry, then 
\begin{equation}
\left|\langle n|\alpha\rangle\right|^{2}=\left|\langle n+k|\alpha\rangle\right|^{2}=\text{const.}
\end{equation}
and 
\begin{equation}
\sum_{n}X_{n}^{\mu\alpha}X_{n}^{\alpha\nu}=\left|\langle n|\alpha\rangle\right|^{2}\delta_{\mu\nu}.\label{eq:XX}
\end{equation}
Therefore, all the off-diagonal terms are zero.

Usually, real systems do not have perfect translational symmetries,
but have some defects or static disorders. In such cases the system
can be described by $H_{0}$, which has perfect translational symmetry,
plus $\delta V$, which breaks this symmetry. If $\delta V$ can be
treated as a perturbation, it is easy to show that the off-diagonal
terms of $\mathbf{K}$ is of order $O\left(\delta V^{4}\right)$,
and can be omitted safely.

\subsection{Reliability of the cumulant expansion for the emission spectrum}

The emission spectra shown in Fig.~\ref{fig:AE_DeLocal} indicate
that the cumulant expansion can be problematic when the donor is highly
delocalized. However, the cumulant expansion of the absorption spectra
is still quite reliable in this case. The most significant difference
between the emission and absorption spectra lies in the initial states.
For the absorption spectrum, the initial state is factorized and the
bath is Gaussian. This Gaussian property is captured quite well by
the 2nd-order cumulant expansion. For the emission spectrum, the initial
state is entangled and the bath is non-Gaussian. The deviation of
the donor's bath from a Gaussian bath is determined by the reorganization
energy, which can be viewed as a displacement to the bath.

According to the Hamiltonian (\ref{H_sb}), the bath operator couples
with the donor's site operator independently. Therefore, when the
donor is highly localized, approximately, each bath operator is displaced
by a scalar reorganization energy. After this displacement the bath
is still Gaussian and the cumulant expansion is safe. However, if
the donor is highly delocalized, the displacement is not a scalar
any more, and the bath is not Gaussian. This problem becomes serious
when the donor's energy gap is larger than the thermal energy, i.e.
$\left|\beta H_{s}^{D}\right|>1$. In this case, the cumulant expansion
of the imaginary-time part is unreliable. We should note that this
is the case of the LH2, even when $T=300$K.

Below, we give a concrete example to discuss this problem. We consider
a fully delocalized donor, 
\begin{equation}
H_{s}^{D}=\begin{pmatrix}0 & V\\
V & 0
\end{pmatrix}.\label{eq:donor_Ham_sym}
\end{equation}
For this system, as we just showed in the previous section the matrices
$\mathbf{K}^{II}$, $\mathbf{K}^{RR}$ and $\mathbf{K}^{RI}$ are
diagonal in the energy representation.

The matrix $\mathbf{K}^{RR}$ is obtained from the 2nd-order cumulant
expansion of the real-time part. It depends on both the time and the
temperature. According to Eqs.~(\ref{eq:KRR}) and (\ref{eq:XX}),
since the donor's Hamiltonian (\ref{eq:donor_Ham_sym}) here has translational
symmetry, $\mathbf{K}^{RR}$ is diagonal, and it does not depend on
the temperature. This term should be reliable since we obtain very
accurate absorption spectra as shown in Figs.~\ref{fig:AE_Local}
and \ref{fig:AE_DeLocal}. $\mathbf{K}^{II}$ comes from the 2nd-order
correction of the equilibrium state, and is unreliable for low-temperature
case.

The matrix $\mathbf{K}^{RI}$ comes from the first-order correction
of the real-time part and the first-order correction of the imaginary-time
(temperature) part. It is diagonal when we use the Hamiltonian (\ref{eq:donor_Ham_sym}),
and the diagonal elements are 
\begin{eqnarray}
K_{\mu\mu}^{RI} & \simeq & i\frac{2\lambda}{\beta}e^{-\gamma t}\sum_{\alpha=1}^{N_{D}}\frac{e^{i\omega_{\mu\alpha}\left(t-i\beta\right)}}{\lambda^{2}+\omega_{\mu\alpha}^{2}},\label{eq:K_RI_mm}
\end{eqnarray}
where the Drude spectrum (\ref{eq:Drude_Spectrum}) and the high-temperature
limit $\cot\frac{\beta\gamma}{2}\simeq\frac{2}{\beta\gamma}$ are
used (see Appendix D). From the above expression we see that all the
excited states $|\alpha\rangle$ will contribute to the matrix element
$K_{\mu\mu}^{RI}$. 
\begin{figure}
\includegraphics[width=1\columnwidth]{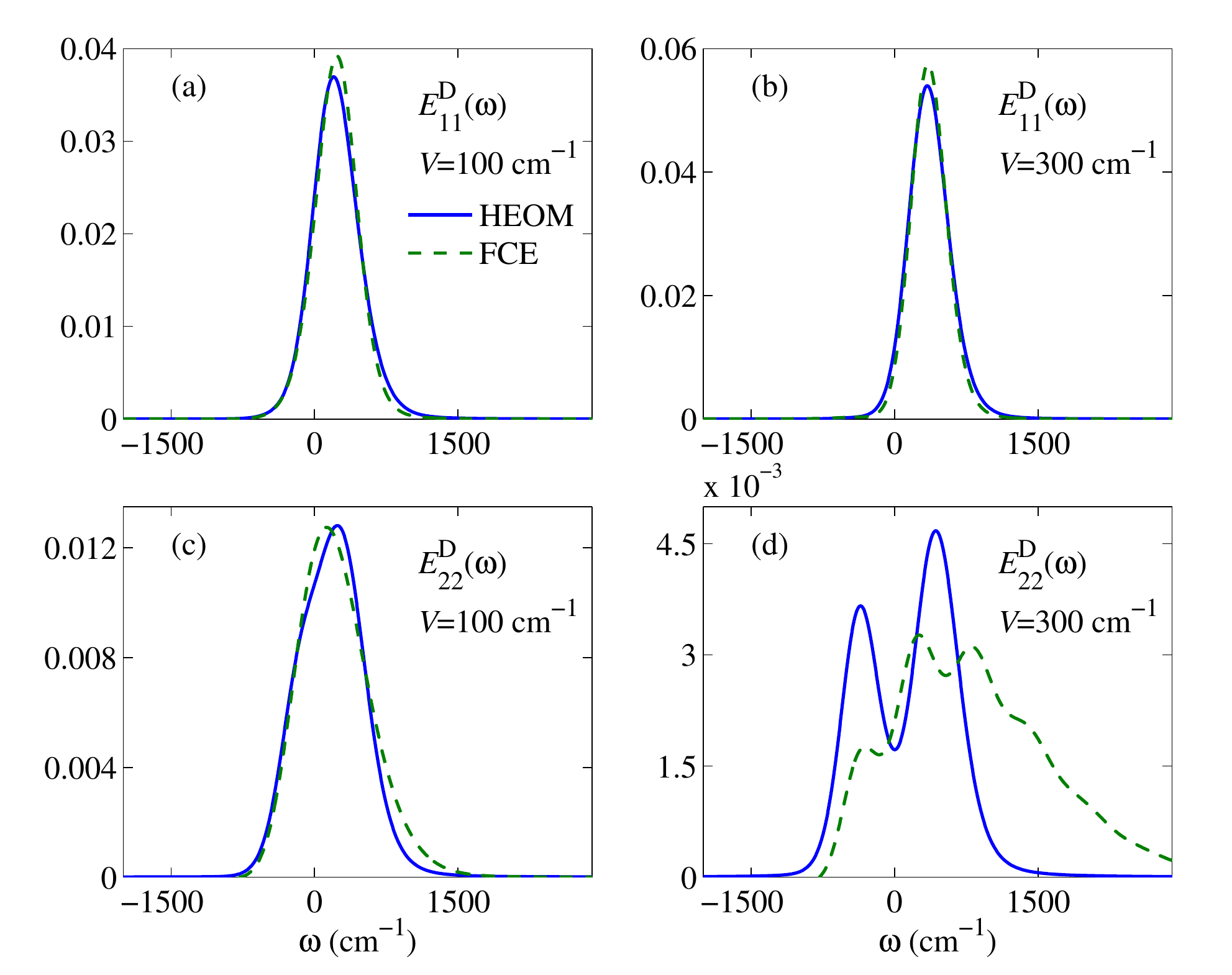}

\protect\caption{Comparison of the emission spectra obtained by the hierarchy equation
of motion (HEOM) and the full 2nd-order cumulant expansion (FCE) methods.
The reorganization energy is $\lambda=200\text{cm}^{-1}$.}

\label{fig:Emission_Spec_Compare} 
\end{figure}

The summation of $\alpha$ in Eq.~(\ref{eq:K_RI_mm}) can be divided
into two parts: (i) $\mu>\alpha$, and thus $\exp\left(\beta\omega_{\mu\alpha}\right)>1$.
(ii) $\mu\le\alpha$, and thus $\exp\left(\beta\omega_{\mu\alpha}\right)\le1$.
If $|\mu\rangle$ is a low-excitated state, we have $\exp\left(\beta\omega_{\mu\alpha}\right)\le1$
for most $\alpha$, and $K_{\mu\mu}^{RI}$ will not become a very
large value. On the opposite side, if $|\mu\rangle$ is a high-excitated
state, $\omega_{\mu\alpha}$ could be a very large positive value
and $\exp\left(\beta\omega_{\mu\alpha}\right)\gg1$. In this case
the matrix element $K_{\mu\mu}^{RI}$ could result in an unreliable
dynamics of $E_{\mu\mu}^{D}\left(t\right)$.

Consider the Hamiltonian (\ref{eq:donor_Ham_sym}), we can obtain
\begin{eqnarray}
K_{11}^{RI} & \simeq & i\frac{2\lambda e^{-\gamma t}}{\beta}\left(\frac{1}{\lambda^{2}}+\frac{e^{-2iVt}}{\lambda^{2}+\omega_{\mu\alpha}^{2}}e^{-2\beta V}\right),\nonumber \\
K_{22}^{RI} & \simeq & i\frac{2\lambda e^{-\gamma t}}{\beta}\left(\frac{1}{\lambda^{2}}+\frac{e^{2iVt}}{\lambda^{2}+\omega_{\mu\alpha}^{2}}e^{2\beta V}\right),
\end{eqnarray}
where $K_{22}^{RI}$ contains a term that diverges as $\exp\left(2\beta V\right)$.
In the energy representation, since all the matrices $\mathbf{K}^{II}$,
$\mathbf{K}^{RR}$ and $\mathbf{K}^{RI}$ are diagonal, the emission
spectrum matrix $\mathbf{E}^{D}\left(\omega\right)=\sum_{\mu}E_{\mu\mu}^{D}\left(\omega\right)|\mu\rangle\langle\mu|$
is also diagonal. In Fig.~\ref{fig:Emission_Spec_Compare}, we show
the deviation of the emission spectra $E_{\mu\mu}^{D}\left(\omega\right)$
obtained by the cumulant expansion and the HEOM for different off-diagonal
coupling $V$. The upper two panels show the emission spectrum of
the lower excitation level $\mu=1$, for which the precision of the
spectrum obtained by the cumulant expansion method is very reliable
for different off-diagonal coupling $V$. However, from the lower
panels of Fig.~\ref{fig:Emission_Spec_Compare}, the spectrum $E_{22}^{D}\left(\omega\right)$
obtained from the cumulant expansion deviates from the exact one as
the increase of $V$.

Therefore, if the emission spectrum is determined mainly by the excited
states that below the thermal energy, the cumulant expansion method
is still reliable. This is the case when we calculate some far-field
emission spectra, where the system's dipole operators will select
the lowest excited state.

\subsection{Multichromophoric FRET Rate}

\begin{figure}
\includegraphics[width=1\columnwidth]{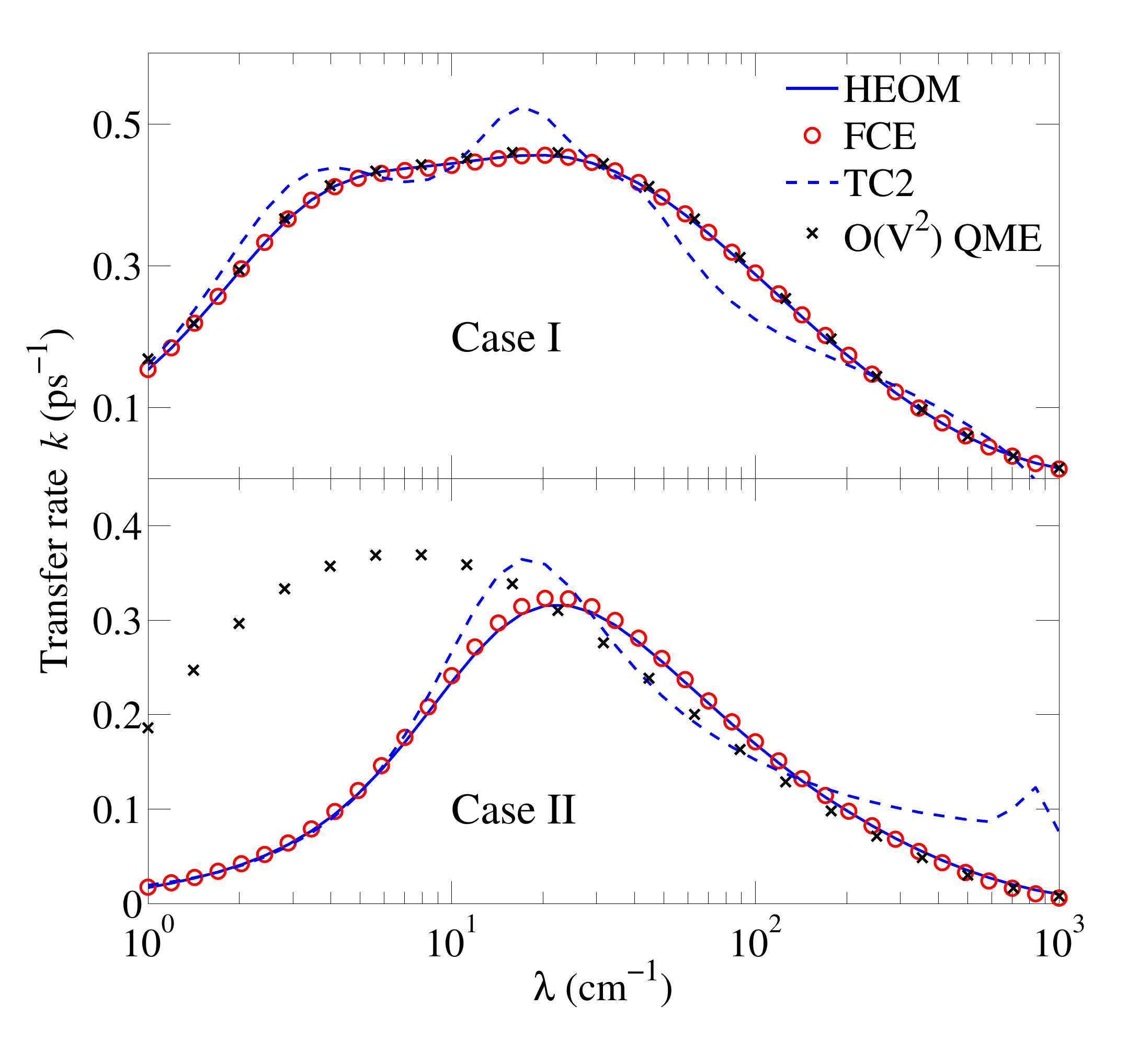}

\protect\caption{Comparison of multichromophoric FRET rates obtained by hierarchy equation
of motion (HEOM), full 2nd-order cumulant expansion (FCE), TC2 \cite{jang_multichromophoric_2004}
and $O\left(V^{2}\right)$ quantum master equation (QME) methods,
as a function of reorganization energy $\lambda$ for localized (\ref{eq:Ham_Local})
and delocalized (\ref{eq:Ham_DeLocal}) cases. The bath parameters
are the same as in Fig.~\ref{fig:comp_Emission_spectra}.}

\label{fig:MC-FRET_Compare} 
\end{figure}

After the study of spectra, we can calculate the multichromophoric
FRET rate. In Fig.~\ref{fig:MC-FRET_Compare}, we compare the multichromophoric
FRET rate obtained via different methods. The exact results are obtained
by HEOM. In this paper, the cumulant expansion is performed with respect
to $\lambda$. We can also do perturbation with respect to the inter-site
coupling $V$ of Eq.(\ref{eq:Donors_Hamiltonian}). This approach
\cite{Cleary} has a precision of $O\left(V^{2}\right)$, as shown
in Fig.~\ref{fig:MC-FRET_Compare}. For a localized system, $V$
is a good perturbative parameter, while for a delocalized system this
method can give reliable results only for $\lambda\gg V$. The rate
given by the TC2 method is qualitatively tolerant. It is not stable
for very large reorganization energy.

Although the emission spectra could be not very precise for delocalized
systems, the rate obtained by our cumulant expansion method is still
in very good agreement with the exact one, since the multichromophoric
FRET rate is proportional to the overlap integral between emission
and absorption spectra. If the reorganization energy $\lambda$ is
very large, the height of the spectra is very low and the overlap
of the spectra is small.

\section{Conclusion}

In this paper, we study the multichromophoric FRET rate and the spectra,
based on a full 2nd-order cumulant expansion in both real- and imaginary-time
domains, which treats the entire system-bath interaction Hamiltonian
perturbatively, and can reduce to the exact FRET for monomers.

(i) In the emission spectrum, the initial state is an equilibrium
state of the donor and its bath. Due to their interaction, both the
donor and the bath deviate from their Boltzmann distributions. Moreover,
the equilibrium state cannot be written in a factorized form, and
the entanglement between the donor and its bath will affect the subsequent
real-time dynamics. The failure of factorization approaches shows
the crucial role of the donor-bath entanglement in both the emission
spectrum and the multichromophoric FRET rate. 

(ii) The FCE method is applied in both localized and delocalized systems.
The absorption spectra obtained by the FCE method are in very good
agreement with the exact results for both localized and delocalized
cases. Further approximations of the FCE can give the IPR and the
OCE methods, which overlook the importance of the off-diagonal system-bath
coupling and fail to give reliable absorption spectra when the system
is highly delocalized.

(iii) The calculation of the emission spectrum is more complicated
due to the initial donor-bath entanglement, which depends on the donor-bath
interaction and the degree of delocalization. For localized system,
the entanglement is weak, and FCE method performs well. For delocalized
system, the FCE method can still give reliable results for low-excitation
state, while the method becomes unreliable for high-excitation states
in the strong system-bath coupling regime. This problem is partially
solved in our Paper II by combining the cumulant expansion with imaginary-time
path integrals.

(iv) In contrast with the spectra, the multichromophoric FRET rate
is more robust since it is proportional to the integral overlap between
the emission and absorption spectra. The deviations in spectra are
reduced in the transfer rate. Moreover, if the reorganization energy
$\lambda$ is very large, the height of the spectra is very low and
the overlap of the spectra is small. Thus, although the emission spectrum
obtained by cumulant expansion could be not very reliable in strong
system-bath coupling regime, we can still obtain a good enough transfer
rate.

(v) The FCE method cannot give reliable emission spectra of delocalized
systems, when the reorganization energy is large and the thermal energy
is small. We develop several new methods to overcome this problem.
When the reorganization energy $\lambda$ is dominate, perturbation
can be carried out for system's off-diagonal coupling $V$ up to its
2nd-order \cite{Cleary}. For more complicated systems such as the
LH2, traditional perturbation methods fail to give reliable emission
spectrum and multichromophoric FRET rate, since the energy gap of
the first excitations, the thermal energy, and the reorganization
energy are comparable. In our new developed hybrid cumulant expansion
method, we use the imaginary-time path integrals to obtain the exact
reduced density matrix of the donor, from which the displacements
of the bath operators can be extracted more precisely. This hybrid
method can give much more reliable emission spectrum and multichromophoric
FRET rate for systems like LH2. Furthermore, to overcome the problems
of the HEOM method in calculating large system and low-temperature
conditions, we implement a complex-time stochastic path integrals
method \cite{Moix2013_PI_FRET}, which gives us the benchmark.
\begin{acknowledgments}
This work was supported by the National Science Foundation (Grant
CHE-1112825) and the Defense Advanced Research Planning Agency (Grant
N99001-10-1-4063). Jian Ma also thanks Liam Cleary and Jeremy Moix
for helpful discussions. 
\end{acknowledgments}
\appendix

\begin{widetext}

\section{Bath correlation function and Lineshape matrices}

\subsection{Bath correlation function}

The general form of the bath correlation function can be derived as

\begin{eqnarray}
C^{B}\!\left(t\!-\! i\tau\right) & \!\!\!= & \!\!\!\int_{0}^{\infty}\frac{d\omega}{\pi}J\left(\omega\right)\frac{\cosh\left[\omega\left(\frac{1}{2}\beta-i\left(t-i\tau\right)\right)\right]}{\sinh\left[\frac{1}{2}\omega\beta\right]}\nonumber \\
 & \!\!\!=\!\!\! & \frac{4\lambda}{\beta}\bigg\{\frac{1}{2}e^{-\gamma\left|t\right|}\!\!+\!\!\gamma\sum_{k=1}^{\infty}\frac{\cos\left(\nu_{k}\tau\right)\left(\gamma e^{-\gamma\left|t\right|}\!\!-\!\!\nu_{k}e^{-\nu_{k}\left|t\right|}\right)}{\gamma^{2}-\nu_{k}^{2}}\!\!-\!\! i\text{sgn}\left(t\right)\gamma\sum_{k=1}^{\infty}\frac{\sin\left(\nu_{k}\tau\right)\left(\nu_{k}e^{-\gamma\left|t\right|}\!\!-\!\!\nu_{k}e^{-\nu_{k}\left|t\right|}\right)}{\gamma^{2}-\nu_{k}^{2}}\bigg\},
\end{eqnarray}
where $J\left(\omega\right)$ is the Drude spectrum, $\nu_{k}=2\pi k/\beta$
is the Matsubara frequency, and $\text{sgn}\left(x\right)$ is the
sign function.

\subsection{Lineshape matrix $\mathbf{K}\left(t\right)$ }

The matrix $\mathbf{K}\left(t\right)$ in Eq.~(\ref{eq:K}) is given
by 
\begin{eqnarray}
\mathbf{K}\left(t\right) & = & \!\int_{0}^{t}\!\! dt_{2}\int_{0}^{t_{2}}\!\!\! dt_{1}\text{tr}_{b}\left[H_{sb}^{A}\left(t_{2}\right)H_{sb}^{A}\left(t_{1}\right)\rho_{b}^{A}\right]\nonumber \\
 & = & \sum_{\mu,\nu=1}^{N_{A}}|\mu\rangle\langle\nu|\sum_{\alpha=1}^{N_{A}}\sum_{n=1}^{N_{A}}X_{n}^{\mu\alpha}X_{n}^{\alpha\nu}\!\int_{0}^{t}\!\! dt_{2}\int_{0}^{t_{2}}\!\!\! dt_{1}e^{i\omega_{\mu\alpha}t_{2}-i\omega_{\nu\alpha}t_{1}}C^{B}\left(t_{2}\!-t_{1}\!\right),
\end{eqnarray}
where the bath correlation function is 
\begin{eqnarray*}
C^{B}\left(t_{2}-t_{1}\right) & = & \lambda\gamma\left[\cot\left(\frac{\gamma\beta}{2}\right)-i\right]e^{-\gamma\left|t_{2}-t_{1}\right|}+\frac{4\lambda\gamma}{\beta}\sum_{n=1}^{\infty}\frac{\nu_{n}e^{-\nu_{n}\left|t_{2}-t_{1}\right|}}{\nu_{n}^{2}-\gamma^{2}},
\end{eqnarray*}
In the high-temperature limit we can neglect all the Matsubara terms,
and thus 
\begin{equation}
C^{B}\left(t_{2}-t_{1}\right)=\lambda\gamma\left[\cot\left(\frac{\gamma\beta}{2}\right)-i\right]e^{-\gamma\left|t_{2}-t_{1}\right|}.
\end{equation}
In this case, the matrix elements can be derived as 
\begin{eqnarray*}
K_{\mu\nu}\left(t\right) & = & \sum_{\alpha}\sum_{n}X_{n}^{\mu\alpha}X_{n}^{\alpha\nu}\int_{0}^{t}dt_{2}\int_{0}^{t_{2}}dt_{1}e^{i\omega_{\mu\alpha}t_{2}-i\omega_{\nu\alpha}t_{1}}C\left(t_{2}-t_{1}\right)\\
 & = & \sum_{\alpha}\sum_{n}X_{n}^{\mu\alpha}X_{n}^{\alpha\nu}\lambda_{n}\gamma\left[\cot\left(\frac{\gamma\beta}{2}\right)-i\right]F_{\mu\alpha\nu}\left(t\right),
\end{eqnarray*}
where 
\begin{equation}
F_{\mu\alpha\nu}\left(t\right)=\frac{e^{-\gamma t+i\omega_{\mu\alpha}t}-1}{\left(\gamma-i\omega_{\mu\alpha}\right)\left(\gamma-i\omega_{\nu\alpha}\right)}+\frac{e^{i\omega_{\mu\nu}t}-1}{i\omega_{\mu\nu}\left(\gamma-i\omega_{\nu\alpha}\right)}.
\end{equation}
If $\mu=\nu$, we have 
\begin{eqnarray}
F_{\mu\mu\mu} & = & \frac{e^{-\gamma t}-1}{\gamma^{2}}+\frac{t}{\gamma}.
\end{eqnarray}

\subsection{Lineshape matrix $\mathbf{K}^{II}\left(\beta\right)$}

The matrix $\mathbf{K}^{II}\left(\beta\right)$ in Eq.~(\ref{eq:KII})
is 
\begin{eqnarray*}
\mathbf{K}^{II}\left(\beta\right) & = & \int_{0}^{\beta}\!\! d\tau_{2}\int_{0}^{\tau_{2}}\!\!\! d\tau_{1}\text{tr}_{b}\left[H_{sb}^{D}\!\!\left(-i\tau_{2}\right)H_{sb}^{D}\!\!\left(-i\tau_{1}\right)\rho_{b}^{D}\right]\\
 & = & \sum_{\mu\nu\alpha}\sum_{n}X_{n}^{\mu\alpha}X_{n}^{\alpha\nu}|\mu\rangle\langle\nu|\int_{0}^{\beta}d\tau'e^{\omega_{\mu\nu}\tau'}\int_{0}^{\tau'}d\tau\, e^{\omega_{\nu\alpha}\tau}C^{B}\left(-i\tau\right)
\end{eqnarray*}
where the imaginary-time correlation function is 
\begin{eqnarray}
C^{B}\left(-i\tau\right) & = & \frac{2\lambda}{\beta}+\frac{4}{\beta}\sum_{k=1}^{\infty}\frac{\lambda\gamma}{\gamma+\nu_{k}}\cos\left(\nu_{k}\tau\right).
\end{eqnarray}
Substituting the above result into $\mathbf{K}^{II}$, we can solve
the integral 
\begin{equation}
\int_{0}^{\tau'}d\tau\, e^{\omega_{\nu\alpha}\tau}C^{B}\left(-i\tau\right)=\frac{2\lambda}{\beta}\mathcal{F}_{\nu\alpha},
\end{equation}
where 
\begin{eqnarray}
\mathcal{F}_{\nu\nu} & = & \tau'+2\gamma\sum_{k=1}^{\infty}\frac{1}{\gamma+\nu_{k}}\frac{\sin\left(\nu_{k}\tau'\right)}{\nu_{k}},\nonumber \\
\mathcal{F}_{\nu\alpha} & = & \frac{e^{\omega_{\nu\alpha}\tau'}-1}{\omega_{\nu\alpha}}+2\gamma\sum_{k=1}^{\infty}\frac{e^{\omega_{\nu\alpha}\tau'}\left[\nu_{k}\sin\left(\nu_{k}\tau'\right)+\omega_{\nu\alpha}\cos\left(\nu_{k}\tau'\right)\right]-\omega_{\nu\alpha}}{\left(\gamma+\nu_{k}\right)\left(\nu_{k}^{2}+\omega_{\nu\alpha}^{2}\right)}.
\end{eqnarray}

\subsection{Lineshape matrix $\mathbf{K}^{RR}\left(t,\beta\right)$}

The matrix $\mathbf{K}^{RR}\left(t,\beta\right)$ in Eq.~(\ref{eq:KRR})
is

\begin{eqnarray*}
\mathbf{K}^{RR}\left(t,\beta\right) & = & \!\!\int_{0}^{t}\!\!\! ds_{2}\int_{0}^{s_{2}}\!\!\! ds_{1}\text{tr}_{b}\left[H_{sb}^{D}\!\!\left(s_{2}\!\!-\!\! i\beta\right)H_{sb}^{D}\!\!\left(s_{1}\!\!-\!\! i\beta\right)\rho_{b}^{D}\right]\\
 & = & \sum_{\mu\nu\alpha}\sum_{n}X_{n}^{\mu\alpha}X_{n}^{\alpha\nu}|\mu\rangle\langle\nu|e^{\beta\omega_{\mu\nu}}\int_{0}^{t}\!\!\! ds_{2}\, e^{i\omega_{\mu\nu}s_{2}}\int_{0}^{s_{2}}\!\!\!\! ds_{1}\, e^{i\omega_{\nu\alpha}s_{1}}\, C^{B}\left(s_{1}\right),
\end{eqnarray*}
where 
\begin{eqnarray}
C^{B}\left(s\right) & \simeq & \lambda\gamma\left[\cot\left(\frac{\gamma\beta}{2}\right)-i\right]e^{-\gamma s}.
\end{eqnarray}

\subsection{Lineshape matrix $\mathbf{K}^{RI}\left(t,\beta\right)$}

The matrix $\mathbf{K}^{RI}\left(t,\beta\right)$ in Eq.~(\ref{eq:KRI})
is 
\begin{eqnarray*}
\mathbf{K}^{RI}\left(t,\beta\right) & = & \!\!\int_{0}^{t}\!\!\! ds\int_{0}^{\beta}\!\!\! d\tau\text{tr}_{b}\left[H_{sb}^{D}\!\!\left(s\!-\! i\beta\right)H_{sb}^{D}\!\!\left(-i\tau\right)\rho_{b}^{D}\right]\\
 & = & \sum_{\mu\nu\alpha}\sum_{n}X_{n}^{\mu\alpha}X_{n}^{\alpha\nu}|\mu\rangle\langle\nu|e^{\beta\omega_{\mu\alpha}}\int_{0}^{t}\!\!\! ds\int_{0}^{\beta}\!\!\!\, d\tau e^{i\omega_{\mu\alpha}s-\omega_{\nu\alpha}\tau}C^{B}\!\left(-s\!-\! i\tau\right),
\end{eqnarray*}
where 
\begin{eqnarray}
C^{B}\!\left(-s\!-\! i\tau\right) & = & \!\!\!\frac{4\lambda}{\beta}\bigg\{\frac{1}{2}e^{-\gamma s}+\gamma\sum_{k=1}^{\infty}\frac{\cos\left(\nu_{k}\tau\right)\left(\gamma e^{-\gamma s}-\nu_{k}e^{-\nu_{k}\tau}\right)}{\gamma^{2}-\nu_{k}^{2}}\nonumber \\
 &  & +i\gamma\sum_{k=1}^{\infty}\frac{\sin\left(\nu_{k}\tau\right)\left(\nu_{k}e^{-\gamma s}-\nu_{k}e^{-\nu_{k}s}\right)}{\gamma^{2}-\nu_{k}^{2}}\bigg\}.
\end{eqnarray}

\end{widetext}


%

\end{document}